\documentclass[conference,10pt,a4paper]{IEEEtran}
\usepackage[cmex10]{amsmath}
\usepackage{amsfonts}
\usepackage{algorithmic,array,blackboard,cite,mdwmath,mdwtab,eqparbox}
\ifCLASSINFOpdf
  \usepackage[pdftex]{graphicx}
  \graphicspath{{figures/pdf/}{figures/png/}{figures/jpg/}}
  \DeclareGraphicsExtensions{.pdf,.jng,.jpeg}
\else
  \usepackage[dvips]{graphicx}
  \graphicspath{{figures/eps2/}}
  \DeclareGraphicsExtensions{.eps}
\fi
\usepackage[caption=false,font=footnotesize]{subfig}
\usepackage{fixltx2e}
\usepackage{url}
\usepackage{tikz}
\newcounter{exNo}
\def\vec#1{{\bf#1}}
\def\op#1{\hat{#1}}
\def\op#1{#1}
\def\ket#1{| #1 \rangle}

\def\norm#1{\| #1 \|}

\def\Tr{\operatorname{Tr}}

\font\liouvop=eufm10
\def\GG{{\text{\liouvop G}}}

\def\D{\mathcal{D}}
\def\F{\mathcal{F}}

\def\H{\mathcal{H}}
\def\L{\mathcal{L}}

\def\sz{\op{\sigma}_z}

\def\Had{\mathop{\rm Had}}
\def\CNOT{\mathop{\rm CNOT}}

\def\I{\mathrm{I}}
\def\X{\mathrm{X}}
\def\Y{\mathrm{Y}}
\def\Z{\mathrm{Z}}
\def\T{\mathrm{T}}
\def\EPSRC{\mbox{EPSRC}}
\def\QIPIRC{\mbox{QIP IRC}}

\begin{document}
\title{Control Paradigms for Quantum Engineering}
\author{\IEEEauthorblockN{\underline{Sonia G Schirmer}\IEEEauthorrefmark{1},
Gajendran Kandasamy\IEEEauthorrefmark{2} and Simon J. Devitt\IEEEauthorrefmark{3}}
\IEEEauthorblockA{\IEEEauthorrefmark{1} 
                  Department of Applied Maths \& Theoretical Physics, Univ. of Cambridge,\\ 
                  Wilberforce Rd, Cambridge, CB3 0WA, UK;  
                  Email: sgs29@cam.ac.uk}
\IEEEauthorblockA{\IEEEauthorrefmark{2} I\&E Group, Tanaka Business School, 
                  Imperial College London, London SW7 2AZ, UK}
\IEEEauthorblockA{\IEEEauthorrefmark{3} 
Quantum Information Division, National Institute of Informatics,
2-1-2 Hitosubashi, Chiyoda-ku, Tokyo, Japan}}
\IEEEspecialpapernotice{(Invited Paper)}
\maketitle

\begin{abstract}
We give an overview of different paradigms for control of quantum 
systems and their applications, illustrated with specific examples.  
We further discuss the implications of fault-tolerance requirements
for quantum process engineering using optimal control, and explore 
the possibilities for architecture simplification and effective
control using a minimum number of simple switch actuators.   
\end{abstract}

\section{Introduction}

The rapid advancement of nanotechnology and nano-engineering is
creating unprecedented new possibilities, not only to create ever
smaller devices, but to probe physical regimes where classical
behaviour gives way to quantum effects such as quantum tunnelling, 
interference, and entanglement, which fundamentally alter the 
system's behaviour.  For instance, the push for ever smaller and
more powerful integrated circuits and computer chips continues
to require the development of ever more sophisticated modelling 
to take quantum effects into account.  But for quantum technology
to truly succeed in creating new and useful applications, physical 
modelling must be augmented by an engineering perspective.  One 
crucial aspect in this is control.

This paper is mostly concerned with control paradigms for quantum 
engineering.  Quantum engineering is a very diverse 
area covering many different types of physical systems and materials,
from a variety of semiconductor nanostructures and superconducting 
materials to neutral atoms, ions, molecules and macroscopic quantum 
states such as Bose-Einstein condensates.  The objectives of control
are equally diverse, ranging from feedback stabilisation with various
applications such as quantum state reduction, optimal quantum
measurements and laser cooling, to coherent manipulation of quantum
states, quantum processes engineering and decoherence control.  Such
a diversity of systems and objectives requires different control 
approaches.

In the following we start with a brief overview of different control 
paradigms relevant for various quantum engineering applications, and
consider specific examples such as indirect control of nuclear spin
dynamics using electron spins as quantum controllers.  In section 3,
we focus on optimal control paradigms for quantum process engineering
in general, and consider the requirements of fault-tolerant circuit 
design and their implications for optimal control, as well as the 
possibilities of architecture simplification and ``minimalist'' control.

\section{Paradigms for Quantum Control}

Most quantum control strategies fall into one of three categories:
feedback control based on feedback of classical information obtained
from (weak) measurements of the system, coherent feedback control
using quantum actuators, and open-loop Hamiltonian (and sometimes
reservoir) engineering.  

\subsection{Open-loop Hamiltonian (and reservoir) engineering}

The conceptually simplest, yet very important type of quantum control 
is open-loop control in the form of Hamiltonian~\cite{qph0602014} and 
sometimes reservoir engineering~\cite{PRL77p4728}.  Considering that 
the evolution of a quantum system is governed by the Schrodinger 
equation (closed systems), or more generally, the quantum Liouville 
equation (with $\hbar=1$):
\begin{equation}
  \dot{\rho}(t) = -i [H,\rho(t)] + \L_D\rho(t)
                \equiv -i (H\rho(t) - \rho(t)H) - \L_D\rho(t),
\end{equation}
where $\rho(t)$ is a positive unit-trace operator (density operator)
acting on a Hilbert space $\H$, which represents the state of the system, 
this approach involves basically engineering a Hamiltonian operator $H$, 
and possibly a (completely positive) superoperator $\L_D$ to achieve a 
desired evolution of the system.  Hamiltonian engineering is usually 
achieved by applying suitable static or dynamic electromagnetic fields 
$\vec{f}(t)$, that coherently interact with the system, thus modifying 
its intrinsic Hamiltonian $H_0\mapsto H[\vec{f}(t)]=H_0+H_C[\vec{f}(t)]$.
$\L_D$ is determined by the system's interaction with its environment 
and can usually (at least for Markovian systems) be written as
$\L_D[\rho(t)]=\sum_k\D[A_k]\rho(t)$, where $A_k$ are Hilbert space 
operators and the dissipative superoperators are~\cite{76Gorini}
\begin{equation}
  \D[A_k] \rho_c(t) = A_k \rho_c(t) A_k^\dagger 
                   - (A_k^\dagger A_k \rho_c(t) + \rho_c(t) A_k^\dagger A_k)/2.
\end{equation}
$\L_D$ can in principle be altered by environmental (reservoir) 
engineering, although this is usually challenging.

Open-loop Hamiltonian (and sometimes reservoir) engineering plays a 
crucial role in many applications including nuclear and electron spin 
engineering in nuclear magnetic resonance (NMR)~\cite{RMP76p1037} and 
electron spin resonance (ESR) applications, control of 
electronic~\cite{JCP114p8820}, vibrational~\cite{CPL290p75}, 
rotational~\cite{PRA69n043401} and translational degrees of freedom 
of molecular systems, atomic vapours, trapped ions, 
Bose Einstein condensates~\cite{PRA60p4875}, control of chemical reactions 
using photonic reagents in quantum chemistry~\cite{JPCA104p4882}, as
well as control of artificial structures such as quantum well and
quantum dots~\cite{PRL91n226804}, cooper-pair boxes~\cite{NAT398p786}, 
and many other systems.

\subsection{Measurement-based feedback control}

\begin{figure}
\centering\scalebox{0.75}{\includegraphics{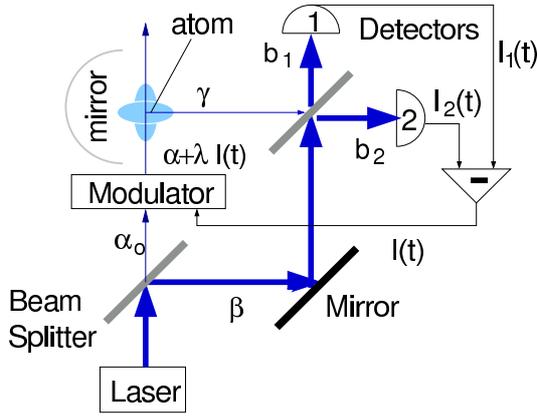}}
\caption{Typical setup for measurement-based feedback control using 
homodyne detection.  The measurement current is fed back to the
field modulator which serves as control actuator.  The model atom can
be an atomic vapour, molecular Bose-Einstein condensate, quantum dot,
etc.}
\label{fig1}
\end{figure}

Another important type of quantum control is measurement-based feedback
control~\cite{PRA49p2133}.  Typically, this approach also involves 
Hamiltonian engineering by applying suitable control fields, but in 
addition the system is monitored, usually via continuous weak
measurements, and the information 
gained from these observations fed back to the actuators as shown in 
Fig.~\ref{fig1}.  Due to the nature of quantum measurements, this 
leads to stochastic evolution governed by a Master equation of the form 
\begin{equation}
  d\rho_c(t) = \left\{-i[H,\rho_c(t)] + \L_D\rho_c(t) \right\} dt
               + \L_M \rho_c(t)\, dW(t),
\end{equation}
where $\rho_c(t)$ is a density operator now representing the conditional 
state (conditioned on the measurement record up to time $t$).  $H$ and 
$\L_D$ are a Hamiltonian and positive superoperator as before, but in 
addition to these deterministic (drift) terms there is now a stochastic 
term $\L_M \rho_c(t) \, dW(t)$, which can usually be written in the
from $\sum_k \H[B_k] \rho_c(t) \, dW(t)$ where
\begin{equation}
  \H[B_k]\rho_c(t) = B_k\rho_c(t)+\rho_c(t) B_k^\dagger
                     -\Tr[B_k\rho_c(t)+\rho_c(t)B_k^\dagger]
\end{equation}
for suitable Hilbert space operators $B_k$, which depend the measurement
operators and feedback Hamiltonian, and $dW(t)$ is the Wiener element of 
the stochastic process.

While open-loop Hamiltonian engineering usually involves control of
non-equilibrium dynamics, often on nano-, pico or femtosecond timescales, 
measurement-based feedback control is very important for control of 
equilibrium dynamics, including steering the system to a steady state%
~\cite{PRA64n063810} with applications in quantum state 
reduction~\cite{qph0402136}, laser cooling of atomic or molecular 
motion~\cite{PRL92n223004}, control of solid-state
qubits~\cite{PRB66n041401}, decoherence control~\cite{qph0603021}
and quantum metrology.

\subsection{Coherent feedback and indirect control}

A third major paradigm for quantum control is coherent feedback control%
\cite{lloyd:022108,nelson:3045}.
Unlike measurement-based feedback control, coherent feedback control
does not (at least not directly) involve any classical actuators or 
measurements.  Rather it relies on indirect control of a target quantum
system $A$ through its coherent interaction with another quantum system
$B$ acting as the controller.  Unlike Hamiltonian/reservoir engineering 
and measurement-based feedback control, which are governed by generally
complicated non-linear control equations, when the system and controller 
are both quantum-mechanical and their interaction is fully coherent, the 
resulting control system is often linear, and can be modelled using 
transfer functions~\cite{IEEETAC48p2107}, albeit with (stochastic) 
operators instead of real vectors representing the state of the system 
and controller.  

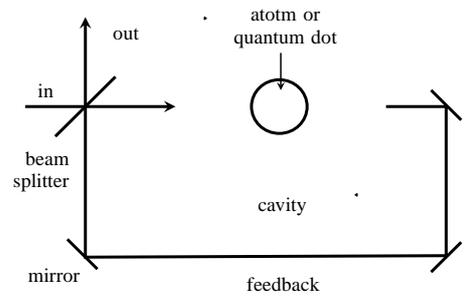
\begin{figure}
\centering\scalebox{0.75}{
\ifx\du\undefined
  \newlength{\du}
\fi
\setlength{\du}{15\unitlength}
\begin{tikzpicture}
\pgftransformxscale{1.000000}
\pgftransformyscale{-1.000000}
\definecolor{dialinecolor}{rgb}{0.000000, 0.000000, 0.000000}
\pgfsetstrokecolor{dialinecolor}
\definecolor{dialinecolor}{rgb}{1.000000, 1.000000, 1.000000}
\pgfsetfillcolor{dialinecolor}
\pgfsetlinewidth{0.100000\du}
\pgfsetdash{}{0pt}
\pgfsetdash{}{0pt}
\pgfsetbuttcap
{
\definecolor{dialinecolor}{rgb}{0.000000, 0.000000, 0.000000}
\pgfsetfillcolor{dialinecolor}
\definecolor{dialinecolor}{rgb}{0.000000, 0.000000, 0.000000}
\pgfsetstrokecolor{dialinecolor}
\pgfpathmoveto{\pgfpoint{7.000126\du}{4.999833\du}}
\pgfpatharc{217}{144}{5.000000\du/5.000000\du}
\pgfusepath{stroke}
}
\pgfsetlinewidth{0.100000\du}
\pgfsetdash{}{0pt}
\pgfsetdash{}{0pt}
\pgfsetbuttcap
{
\definecolor{dialinecolor}{rgb}{0.000000, 0.000000, 0.000000}
\pgfsetfillcolor{dialinecolor}
\definecolor{dialinecolor}{rgb}{0.000000, 0.000000, 0.000000}
\pgfsetstrokecolor{dialinecolor}
\pgfpathmoveto{\pgfpoint{11.999716\du}{11.000378\du}}
\pgfpatharc{37}{-36}{5.000000\du/5.000000\du}
\pgfusepath{stroke}
}
\pgfsetlinewidth{0.100000\du}
\pgfsetdash{}{0pt}
\pgfsetdash{}{0pt}
\pgfsetbuttcap
{
\definecolor{dialinecolor}{rgb}{0.000000, 0.000000, 0.000000}
\pgfsetfillcolor{dialinecolor}
\pgfsetarrowsend{stealth}
\definecolor{dialinecolor}{rgb}{0.000000, 0.000000, 0.000000}
\pgfsetstrokecolor{dialinecolor}
\draw (1.000000\du,8.000000\du)--(6.000000\du,8.000000\du);
}
\pgfsetlinewidth{0.100000\du}
\pgfsetdash{}{0pt}
\pgfsetdash{}{0pt}
\pgfsetmiterjoin
\pgfsetbuttcap
{
\definecolor{dialinecolor}{rgb}{0.000000, 0.000000, 0.000000}
\pgfsetfillcolor{dialinecolor}
\pgfsetarrowsend{stealth}
{\pgfsetcornersarced{\pgfpoint{0.000000\du}{0.000000\du}}\definecolor{dialinecolor}{rgb}{0.000000, 0.000000, 0.000000}
\pgfsetstrokecolor{dialinecolor}
\draw (13.000000\du,8.000000\du)--(15.000000\du,8.000000\du)--(15.000000\du,13.000000\du)--(3.000000\du,13.050000\du)--(3.000000\du,5.000000\du);
}}
\pgfsetlinewidth{0.100000\du}
\pgfsetdash{}{0pt}
\pgfsetdash{}{0pt}
\pgfsetbuttcap
{
\definecolor{dialinecolor}{rgb}{0.000000, 0.000000, 0.000000}
\pgfsetfillcolor{dialinecolor}
\definecolor{dialinecolor}{rgb}{0.000000, 0.000000, 0.000000}
\pgfsetstrokecolor{dialinecolor}
\draw (2.000000\du,9.000000\du)--(4.000000\du,7.000000\du);
}
\pgfsetlinewidth{0.100000\du}
\pgfsetdash{}{0pt}
\pgfsetdash{}{0pt}
\pgfsetbuttcap
{
\definecolor{dialinecolor}{rgb}{0.000000, 0.000000, 0.000000}
\pgfsetfillcolor{dialinecolor}
\definecolor{dialinecolor}{rgb}{0.000000, 0.000000, 0.000000}
\pgfsetstrokecolor{dialinecolor}
\draw (14.500000\du,7.450000\du)--(15.500000\du,8.450000\du);
}
\pgfsetlinewidth{0.100000\du}
\pgfsetdash{}{0pt}
\pgfsetdash{}{0pt}
\pgfsetbuttcap
{
\definecolor{dialinecolor}{rgb}{0.000000, 0.000000, 0.000000}
\pgfsetfillcolor{dialinecolor}
\definecolor{dialinecolor}{rgb}{0.000000, 0.000000, 0.000000}
\pgfsetstrokecolor{dialinecolor}
\draw (2.400000\du,12.450000\du)--(3.400000\du,13.450000\du);
}
\pgfsetlinewidth{0.100000\du}
\pgfsetdash{}{0pt}
\pgfsetdash{}{0pt}
\pgfsetbuttcap
{
\definecolor{dialinecolor}{rgb}{0.000000, 0.000000, 0.000000}
\pgfsetfillcolor{dialinecolor}
\definecolor{dialinecolor}{rgb}{0.000000, 0.000000, 0.000000}
\pgfsetstrokecolor{dialinecolor}
\draw (14.500000\du,13.550000\du)--(15.500000\du,12.550000\du);
}
\definecolor{dialinecolor}{rgb}{1.000000, 1.000000, 1.000000}
\pgfsetfillcolor{dialinecolor}
\pgfpathellipse{\pgfpoint{9.450000\du}{8.000000\du}}{\pgfpoint{0.925000\du}{0\du}}{\pgfpoint{0\du}{0.900000\du}}
\pgfusepath{fill}
\pgfsetlinewidth{0.100000\du}
\pgfsetdash{}{0pt}
\pgfsetdash{}{0pt}
\definecolor{dialinecolor}{rgb}{0.000000, 0.000000, 0.000000}
\pgfsetstrokecolor{dialinecolor}
\pgfpathellipse{\pgfpoint{9.450000\du}{8.000000\du}}{\pgfpoint{0.925000\du}{0\du}}{\pgfpoint{0\du}{0.900000\du}}
\pgfusepath{stroke}
\definecolor{dialinecolor}{rgb}{0.000000, 0.000000, 0.000000}
\pgfsetstrokecolor{dialinecolor}
\node[anchor=west] at (1.150000\du,7.450000\du){in};
\definecolor{dialinecolor}{rgb}{0.000000, 0.000000, 0.000000}
\pgfsetstrokecolor{dialinecolor}
\node[anchor=west] at (3.650000\du,5.550000\du){out};
\definecolor{dialinecolor}{rgb}{0.000000, 0.000000, 0.000000}
\pgfsetstrokecolor{dialinecolor}
\node at (9.650000\du,4.940625\du){atotm or };
\definecolor{dialinecolor}{rgb}{0.000000, 0.000000, 0.000000}
\pgfsetstrokecolor{dialinecolor}
\node at (9.650000\du,5.740625\du){quantum dot};
\definecolor{dialinecolor}{rgb}{0.000000, 0.000000, 0.000000}
\pgfsetstrokecolor{dialinecolor}
\node at (9.550000\du,11.350000\du){cavity};
\definecolor{dialinecolor}{rgb}{0.000000, 0.000000, 0.000000}
\pgfsetstrokecolor{dialinecolor}
\node[anchor=west] at (8.100000\du,13.950000\du){feedback};
\pgfsetlinewidth{0.050000\du}
\pgfsetdash{}{0pt}
\pgfsetdash{}{0pt}
\pgfsetbuttcap
{
\definecolor{dialinecolor}{rgb}{0.000000, 0.000000, 0.000000}
\pgfsetfillcolor{dialinecolor}
\pgfsetarrowsend{stealth}
\definecolor{dialinecolor}{rgb}{0.000000, 0.000000, 0.000000}
\pgfsetstrokecolor{dialinecolor}
\draw (9.500000\du,6.200000\du)--(9.500000\du,7.453223\du);
}
\definecolor{dialinecolor}{rgb}{0.000000, 0.000000, 0.000000}
\pgfsetstrokecolor{dialinecolor}
\node[anchor=east] at (2.700000\du,9.750000\du){beam};
\definecolor{dialinecolor}{rgb}{0.000000, 0.000000, 0.000000}
\pgfsetstrokecolor{dialinecolor}
\node[anchor=east] at (2.700000\du,10.550000\du){splitter};
\definecolor{dialinecolor}{rgb}{0.000000, 0.000000, 0.000000}
\pgfsetstrokecolor{dialinecolor}
\node[anchor=west] at (0.850000\du,13.650000\du){mirror};
\end{tikzpicture}}
\caption{Simple coherent feedback control setup: an atom or quantum dot
interacts with a cavity field; output from the cavity is fed back via
mirrors and a beam splitter.  The input and output fields can be described
by stochastic operators related via a transfer function.}
\label{fig2a}
\end{figure}

\begin{figure}
\centering\scalebox{0.58}{\includegraphics{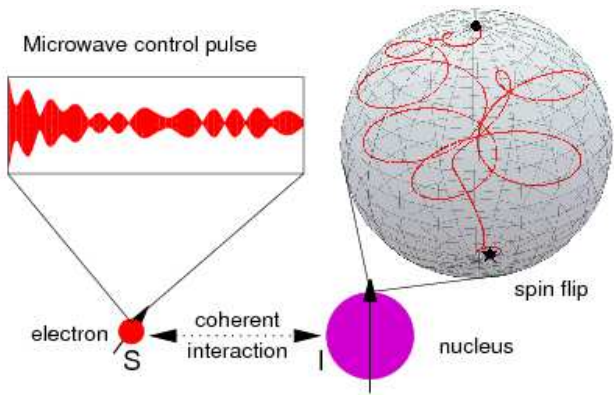}}
\caption{Indirect control of a coupled electron-nuclear spin system: 
A modulated high-frequency control pulse applied to electron spin 
realizes a nuclear spin flip (swap gate).  The trajectory of the Bloch 
vector corresponding to the nuclear spin system shows that the nuclear 
spin initially in the state $\ket{\uparrow}$, indicated by $\bullet$,
is flipped to $\ket{\downarrow}$, indicated by $\ast$.  The specific
system considered here was a 1e1n (one electron, one nuclear spin) 
model of malonic acid described by the Hamiltonian 
$H_0/2\pi = \nu_s S_z + \nu_n I_z + A_{zx} S_z I_z + A_{zz} S_z I_z$,
where $S_k$ and $I_k$ for $k\in\{x,y,z\}$ are the Pauli operators for
the electron and nuclear spin, respectively, and the constants are
$\nu_s=11.885$~GHz, $\nu_n=18.1$~MHz, $A_{zx}=14.2$~MHz, $A_{zz}=-42.7$~MHz.
Indirect control of nuclear spin dynamics using shaped pulses applied to
the electron has been experimentally demonstrated for this 
system~\cite{qph0707-2956}.  The shaped control pulse was calculated
using a varational optimal control approach~\cite{PRA61n012101}.}
\label{fig2b}
\end{figure}

This type of control is interesting in quantum photonics, for example, 
where one can use cavities, mirrors, beam splitters and waveguides to
build optical networks that could control the state of atoms or quantum
dots.  Fig.~\ref{fig2a} shows a very simple coherent feedback system 
involving a cavity and beam splitter.  If an atom or quantum dot is put 
into the cavity, its state could be controlled though this coherent 
feedback.  Another application of coherent feedback is indirect control, 
e.g., the control of nuclear spins by electron spins via the Heisenberg 
interaction, an example of which is shown in Fig.~\ref{fig2b}.

Quantum controllers cannot solve the problem of controlling quantum 
systems completely, however, as the quantum controller itself needs to 
be controlled in some form, and this usually requires interaction with 
a non-quantum system such as classical laboratory equipment at some
stage, and thus control strategies such as Hamiltonian engineering or 
state preparation using measurement-based feedback.  For instance, in
the case of indirect control of nuclear spin dynamics, the electron 
spins need to be controlled using conventional Hamiltonian engineering 
techniques, e.g., by application of tailored control pulses as shown in 
Fig.~\ref{fig2b} to achieve the desired effect on the nuclear spins.  
Indirect control therefore may seem to only complicate the problem---%
considering that nuclear spins can be controlled directly using 
radio-frequency fields, for example.  Yet, the indirect approach is 
promising because exploiting the strength of the quantum interaction 
between the spins and the fact that electron spins can be manipulated 
faster than nuclear spins, enables control of nuclear dynamics on 
shorter timescales than direct control.  Moreover, for some systems
direct control of certain degrees of freedom may not be possible at
all.

\section{Paradigms for Optimal Control Design}

Regardless of the specifics of the system and control paradigm, it is 
always desirable to optimise the control.  But as with control paradigms 
for quantum systems in general, there are many paradigms for optimal 
control.  In many applications in quantum chemistry, optimal control is 
understood to mean finding photonic reagents, typically shaped pulses, 
to maximise the yield of some observable, or reaction product, subject 
to constraints.  Similarly, in quantum information processing the aim
is often to find controls to maximise the fidelity of a quantum process, 
subject to constraints.  In the latter area, it has been argued that 
beside optimising the overall fidelity, time-optimality is crucial, as 
shorter control pulse sequences mean shorter gate operation times, and 
assuming the rate of decoherence is roughly constant, less decoherence, 
and thus higher gate fidelities.  However, some proposed approaches to 
circuit optimisation appear incompatible with fault-tolerant designs,
which in the context of robust, scalable architectures may be a more 
important consideration.  The implementation of non-trivial gates (such 
as Toffoli or T-gates) on encoded logical qubits in fault-tolerant 
designs requires a large number of ancilla qubits, multiple verification 
measurements and classical feed-forward in the standard model, and it 
could be argued that eliminating or reducing some of these cumbersome 
requirements should be an important paradigm for optimal control. 
Another important issue are design constraints.  While a large number 
of sophisticated actuators may be desirable for reasons of flexibility, 
robustness and possibilities for optimisation, in practice, simplicity 
of a design is often a crucial factor, and may be the difference between 
a physically realistic device and a practically infeasible one.

\subsection{Time-optimality vs Fault-tolerance}

An interesting and important application of optimal control has been in 
the area of quantum circuit design.  In the conventional circuit model 
of quantum computing, quantum operations are constructed from a small 
set of elementary gates.  Although any quantum operation on $n$ qubits 
can be constructed from a universal set of gates, a major drawback of 
this approach is that, even if the overhead is polynomial, a very large 
number of gates are required to implement any non-trivial operation, and 
standard sets of elementary gates such as the Hadamard, phase, or CNOT 
gates, are often no easier to implement physically than arbitrary single 
or two-qubit gate.  This approach therefore seems quite inefficient and 
it appears sensible to decompose quantum circuits into larger modules 
that can be implemented efficiently for a particular architecture using 
optimal control theory instead.  It has been shown that the implementation 
time for the quantum Fourier transform circuit, for example, can be 
improved by (at least) a factor of eight in this way~\cite{PRA72n042331}.

A potential problem of this approach, however, is its compatibility with 
fault-tolerant architectures.  No matter how effective the control, 
realistic quantum circuits will require redundancy, error correction and 
fault-tolerant designs to mitigate errors caused not only by imperfect 
control but also by environmental noise and decoherence.  A standard 
approach to fault-tolerant design is encoding information into logical
qubits formed by groups of physical qubits using error correction codes%
\cite{qph9705052}.  
In all existing proposals, fault-tolerant gates on encoded qubits are 
constructed from single and two-qubit logic gates.  It is not clear 
therefore that optimisation of larger circuit blocks acting on physical 
qubits is compatible with fault-tolerant designs.  Furthermore, when 
implementing quantum logic gates on encoded qubits in general, overall 
fidelity and time-optimality are not the only considerations, but we may 
wish to explicitly minimise errors that cannot be corrected vs errors 
that can be corrected for a given encoding.

To illustrate some of the ideas, consider the popular [[7,3,1]] CSS code%
\cite{PRL77p793}, 
which uses seven physical qubits to define one logical qubit, where the 
logical basis states are superpositions of codewords containing an odd 
and even number of ones, respectively,  
\begin{subequations}
\begin{align*}
  \ket{0}_L =& \ket{0000000}+\ket{1111000}+\ket{1100110}+\ket{1010101}\\ 
             &+\ket{0011110}+\ket{0101101}+\ket{0110011}+\ket{1001011}\\
  \ket{1}_L =& \ket{1111111}+\ket{0000111}+\ket{0011001}+\ket{0101010}\\ 
             &+\ket{1100001}+\ket{1010010}+\ket{1001100}+\ket{0110100}.
\end{align*}
\end{subequations}
Assuming we can at least perform measurements of the observables $\X$
and $\Z$ given by the Pauli matrices
\begin{equation}
\label{eq:pauli}
  \X = \begin{pmatrix} 0 & 1 \\ 1 & 0  \end{pmatrix}, \;
  \Y = \begin{pmatrix} 0 & -i \\ i & 0  \end{pmatrix}, \;
  \Z = \begin{pmatrix} 1 & 0 \\ 0 & -1 \end{pmatrix},
\end{equation}
on individual physical qubits, we can extract an error syndrome by
performing parity checks, which project arbitrary states into the $\pm 1$
eigenstates of the stabiliser operators, e.g.,
\begin{multline*}
 \qquad \GG_S = \{\mbox{IIIXXXX},\mbox{IXXIIXX},\mbox{XIXIXIX}, \\
     \mbox{IIIZZZZ},\mbox{IZZIIZZ},\mbox{ZIZIZIZ}\} \qquad\qquad
\end{multline*}
for the CSS code.  On clean code word states errors act to switch the
eigenstate, which can be detected via the parity check and correctly
via local operations on the physical qubits.  The [[7,3,1]] code can 
correct only a single bit or phase flip error.  If more than one error 
occurs then the code fails.  Thus, for the code to be effective, no 
more than one error should occur per error correction block, and a 
single error should not propagate, i.e., if we apply a gate to a state 
with a single error, the output state should not contain more than a 
single error.  

Consider the simplest case of implementing an arbitrary single qubit
gate $U_T$ on a logical qubit, i.e., a unitary operator on $n$ physical 
qubits ($n=7$ for [[7,3,1]] CSS code), in a single step using optimal 
control.  Even if the control is very effective, the operator actually 
implemented will not be exactly $U_T$ but an operator $U_R$ related to 
$U_T$ by an error operator $U_E=U_T^\dagger U_R$, which is the identity 
exactly if $U_R=U_T$.  Since $U_E$ is an operator acting on $n$ physical
qubits, we can expand it with respect to the $n$-qubit Pauli group
$P_n = P^{\otimes n} = \{\I,\X,\Y,\Z\}^{\otimes N}$.  If we define the 
weight of an operator in $P^{\otimes n}$ as the number of non-identity
operations, then the errors we can correct in the standard [[7,3,1]] CSS 
encoding are the contributions of the terms in the expansion that have 
weight $1$, corresponding to single bit or phase flip errors.  Hence,
to maintain a fault-tolerant circuit design we must ensure at least 
$W_1(U_E) \gg W_2(U_E) \gg W_3(U_E)\ldots $, where $W_k(U_E)$ indicates
the contribution of terms with weight $k$ in the Pauli expansion of
$U_E$.  In practice this could be archived by adding constraint terms 
to the functional to be optimised.  For example, instead of maximising 
the (normalised) fidelity $\F(U)=2^{-n}\Tr(U_T^\dagger U)$, one might 
consider maximising $\F(U)-\sum_{k>1}\lambda_k W_k(U_E)$, where 
$\lambda_k$ are suitable weighting factors satisfying 
$\lambda_2\ll \lambda_3\ll \ldots$.  Although preliminary calculations 
for very simple systems suggest that we can find controls that minimise
projections of $U_E$ onto certain subspaces, it is not clear at this 
stage to which extent it is possible to suppress the projections onto 
multi-error subspaces for realistic physical systems, and if circuits 
composed of larger modules implemented using optimal control can be 
made fault-tolerant in general.

\subsection{Architecture Simplification vs Control Sophistication}

Many quantum systems can be controlled by external electromagnetic
fields generated by actuators capable of creating complicated pulses.
For example, optical fields generated by lasers, as well as coherent
microwave or radio-frequency pulses used in ESR and NMR, can be shaped
either using spectral pulse shaping or temporal modulation to create
complex control fields, offering considerable potential for optimisation.
Optimal control in a quantum setting is therefore often understood to
mean optimisation of control pulse shapes.  However, this type of
optimal control is not appropriate in all settings.  An important case
are nano-scale systems such as quantum dots controlled electronically 
using gate electrodes.  Although there is some scope for optimising
voltage profiles, for example, a more important consideration for these
systems is often design optimisation, in particular, minimising the
number of actuators required, and using the simplest actuators that
can accomplish the desired task.  

\begin{figure}
\centering
(a) \scalebox{0.35}{\includegraphics{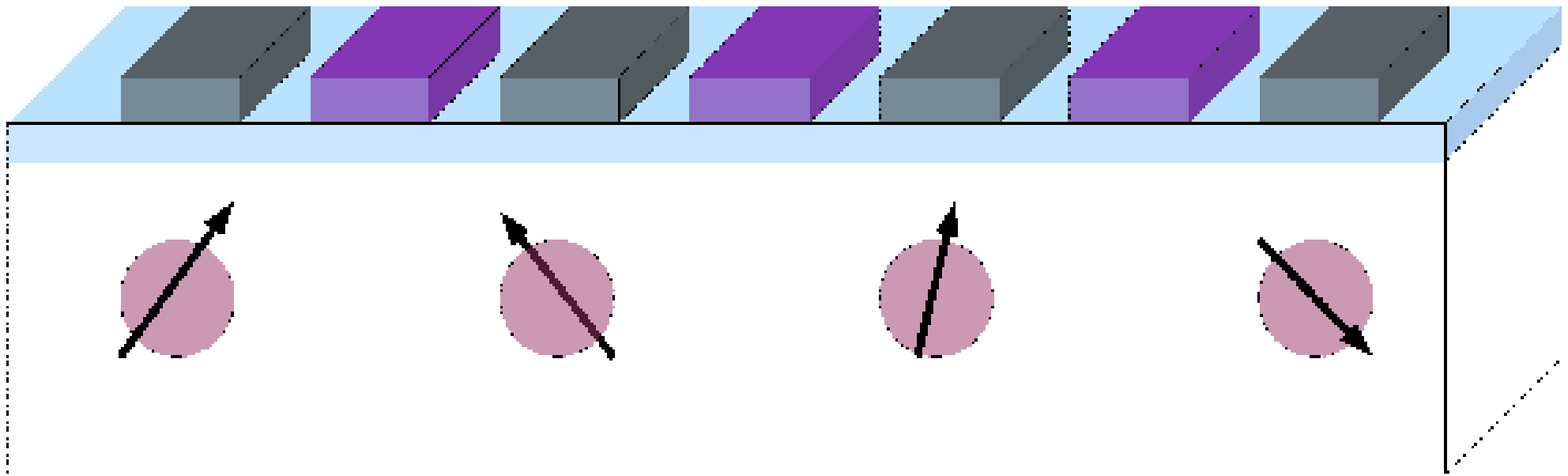}} \\
(b) \scalebox{0.35}{\includegraphics{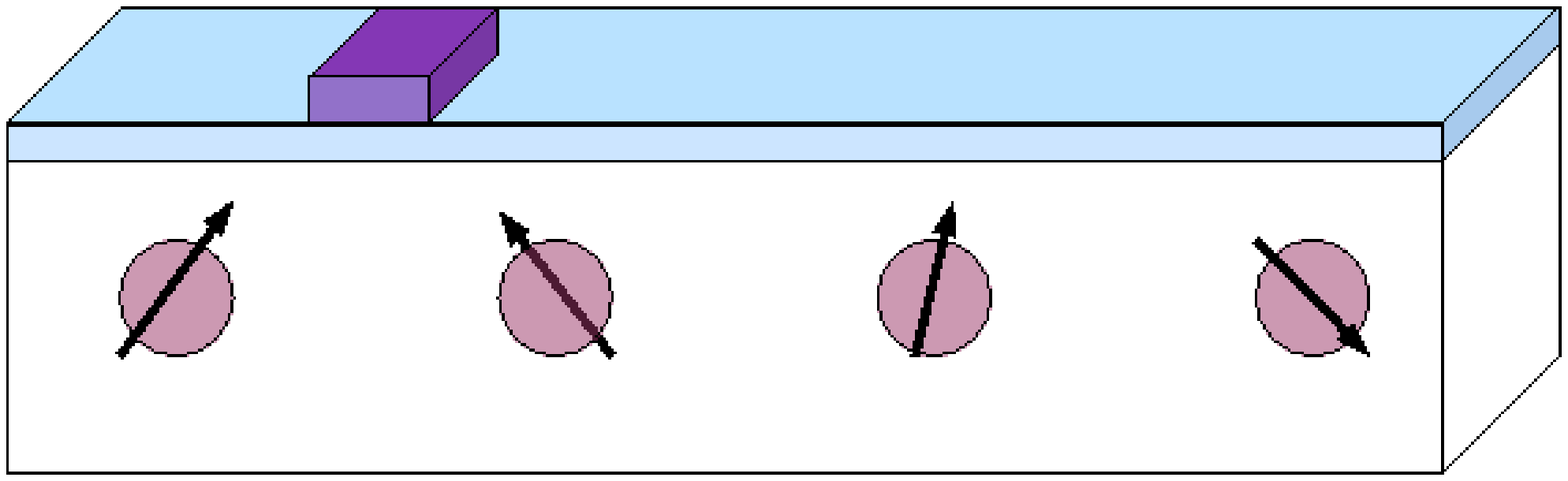}}
\caption{(a) Phosphorus donors embedded in silicon with surface control 
electrodes separated by SiO$_2$ insulating layer (Kane architecture). 
$2n-1$ control electrodes are required to control $n$ spins.  
(b) Spin chain with a single control electrode controlling the entire chain.}
\label{fig3}
\end{figure}

As a specific example, consider the Kane proposal for a solid-state
quantum computer~\cite{NAT393p133}.  The original design involves $2n-1$ 
voltage gates, as pictured in Fig.~\ref{fig3}, to selectively tune $n$ 
quantum dots, in this case the nuclear spins of phosphorus donors in 
silicon, into resonance with a globally applied transverse AC magnetic 
field, and to control the interactions between them.  This design suffers 
from various problems, among them the high density of nanometer-scale
control electrodes required for a scalable architecture, which poses a 
serious challenge for current manufacturing techniques.  Furthermore, 
even if fabrication techniques improve, the presence of a large number 
of closely spaced control electrodes creates fundamental physical 
problems including substantial crosstalk~\cite{NT17p4572} between the 
actuators and the potential for significant decoherence of the quantum 
information stored in the quantum dots via incoherent interaction with 
the control electrodes, which is highly detrimental to device performance.  
Thus, it is highly desirable in this setting to minimise the number of 
control electrodes (i.e., actuators) and to keep the actuators simple.  
A control scheme that requires only binary switch actuators, for example, 
will be advantageous because switching between two fixed voltage settings 
is far simpler than producing complicated temporal voltage profiles, and 
it is easier to experimentally characterise (and compensate) crosstalk 
effects in this setting.  This raises the question of how many of the 
actuators are really necessary.  

Basic controllability analysis of various model systems suggests that
most of the actuators are \emph{not} necessary, and for certain model 
systems it can be shown that a single, local actuator such as a control 
electrode is theoretically sufficient for complete controllability of
the system~\cite{qph0801-0721}.  For instance, a spin chain of length 
$N$ with isotropic Heisenberg interaction decomposes into $N$ excitation 
subspaces with $n$ excitations ($n=1,\ldots,N$).  Assuming nearest 
neighbour coupling, the Hamiltonian of the first excitation subspace is 
tridiagonal
\begin{equation}
 \label{eq:Hspin}
  H_0 = \begin{pmatrix} E_1 & d_1 & 0 & \ldots   & 0 \\
                        d_1 & E_2 & d_2 & \ldots & 0 \\
                        0   & d_2 & E_3 & \ldots & 0 \\
                        \vdots & \vdots & \vdots & \ddots & \vdots\\
                        0   & 0 & 0 & \ldots & E_N 
        \end{pmatrix}
\end{equation}
where $d_n>0$ defines the strength of the interaction between the spins 
$n$ and $n+1$ in the chain for $n=1,\ldots,N-1$ and $E_n$ are the energy
levels.  It can be shown using Lie algebraic techniques that a single 
local actuator that allows us to modulate the coupling between spins $r$
and $r+1$ only, is sufficient in this case for complete controllability
of the \emph{entire} excitation subspace.  This result also holds for 
spin chains with isotropic XY or dipole-dipole coupling.  In fact, the 
controllability result is generic for any type of system with a nearest
neighbour coupling Hamiltonian of the form~(\ref{eq:Hspin}), and similar 
results can be proved for other types of systems.

Although there are many open questions---such as the exact conditions 
for a single actuator to suffice, or the minimum number of actuators 
required for complete controllability for a particular system, these 
preliminary controllability results are encouraging from the point of 
view of architecture simplification---but can we actually find
\emph{constructive} control schemes for systems such as the spin chain
with a single local actuator above? Suppose we have a single actuator 
that has $M$ distinct states.  If the interaction with system is fully
coherent and the system dynamics is therefore Hamiltonian, each of the 
$M$ actuator states is associated with a system Hamiltonians $H_m$ for 
$m=1,\ldots,M$.  Since the only control we have is the ability to switch
between these $M$ Hamiltonians, the unitary operators we can implement 
are of the form
\begin{equation}
\label{eq:Udecomp}
 U(\vec{m},\vec{t}) = U^{(m_1)}(t_1) \ldots 
                      U^{(m_{K-1})}(t_{K-1}) U^{(m_K)}(t_K)
\end{equation}
where $\vec{m}$ and $\vec{t}$ are vectors of length $K$ with 
$m_k\in\{1,\ldots,M\}$ and $t_k\in\RR^+_0$,  and 
$U^{(m_k)}(t_k) = \exp(-i t_k H^{(m_k)})$ are the elementary evolution
operators.  The vector $\vec{m}$ determines the switching sequence and
$\vec{t}$ the switching times, i.e., length of time the system evolves 
under a particular Hamiltonian $H_m$.  The $t_k$ can be regarded as 
generalised Euler angles.  For $M=2$, i.e., a binary switch controller,
Eq.~(\ref{eq:Udecomp}) can be simplified.  Noting that
$U^{(m)}(t_1)U^{(m)}(t_2)=U^{(m)}(t_1+t_2)$, 
we see that the elements of $\vec{m}$ must alternate between $1$ and
$2$, and we have without loss of generality
\begin{equation}
\label{eq:Udecomp2}
 U(\vec{t}) = U^{(1)}(t_1) U^{(2)}(t_2)\ldots 
              U^{(1)}(t_{2\ell-1}) U^{(2)}(t_{2\ell}), 
\end{equation}
if we set $K=2\ell$ and allow the possibility of $t_1=0$ or $t_{\ell}=0$.

In the context of unitary process control, constructive control requires 
therefore that we find vectors $\vec{m}$ and $\vec{t}$ such that
\begin{equation}
  \norm{U_T - U(\vec{m},\vec{t})} < \epsilon
\end{equation}
for a given target operator $U_T$ and tolerance $\epsilon$.  For very
special classes of Hamiltonians, e.g., Hamiltonians that are mutually  
orthogonal with respect to the Hilbert-Schmidt norm, 
$\Tr(H_m^\dagger H_n) \propto\delta_{mn}$, there are various explicit 
decomposition algorithms to compute the switching sequence $\vec{m}$ 
and generalised Euler angles $\vec{t}$.  In practice, we are seldom so
lucky, however.  Indeed for a system with $H_0$ as in Eq.~(\ref{eq:Hspin}) 
and a local binary switch actuator that annuls the coupling between 
spins $r$ and $r+1$, we have $H_r=-d_r (e_{r,r+1}+e_{r+1,r})$, where 
$e_{r,r+1}$ is a matrix that is zero everywhere except for a $1$ in the 
$(r,r+1)$ position, and
\begin{equation}
  \Tr[ H_0^\dagger(H_0+H_r)] = 2 \norm{\vec{d}-\vec{d}_r}^2 + \norm{\vec{E}}^2,
\end{equation}
where $\vec{d}=(d_1,\ldots,d_{N-1})$, $\vec{E}=(E_1,\ldots,E_N)$ and
$\vec{d}_r=(d_{r\ell})$ with $d_{r\ell}=d_\ell\delta_{r\ell}$.  Thus, 
we see that the available Hamiltonians $H_0$ and $H_0+H_r$ corresponding 
to the off and on position of the switch, respectively, are never 
orthogonal, except when all the parameters vanish.  In general
\begin{equation}
  \cos\alpha = \frac{\Tr[H_0^\dagger(H_0+H_r)]}{\Tr[H_0^2] \Tr[(H_0+H_r)^2]}
\end{equation}
is close to $1$ in this case, and thus the angle $\alpha$ between the
Hamiltonians is very small.  Although it is difficult to derive explicit
expressions for the generalised Euler angles and to prove the optimality
of a particular switching sequence, the vectors $\vec{t}$ and $\vec{m}$ 
can be determined numerically using general optimisation techniques. 

\begin{figure}
\centering\scalebox{0.55}{\includegraphics{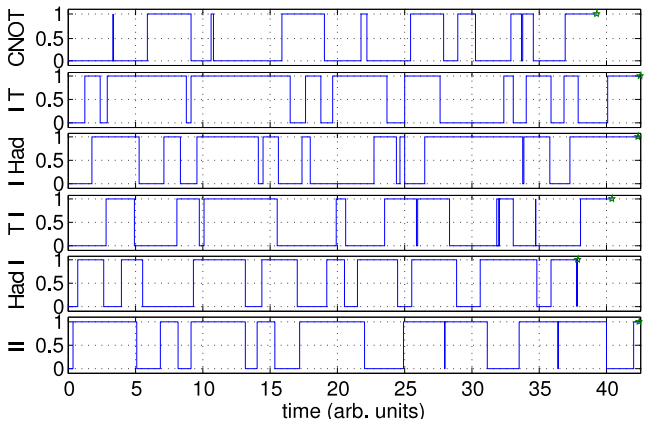}}
\caption{Bang-bang control scheme to implement various elementary gates
on the first excitation subspace of a spin chain of length $4$ with a
single binary switch actuator.  All all pulse sequences have theoretical
fidelities $\ge 99.99$\%.}
\label{fig4}
\end{figure}

As an example, we consider a spin chain of length $4$ with uniform 
isotropic Heisenberg coupling and a single actuator as shown in 
Fig.~\ref{fig3}(b).  Using the natural identification 
\begin{equation*}
 \ket{0}=\ket{00}, \; \ket{1}=\ket{01}, \; 
 \ket{2}=\ket{10}, \; \ket{3}=\ket{11}
\end{equation*} 
of the 1st excitation subspace states, out goal is to implement the six 
``two-qubit'' gates
\begin{equation*}
 \label{eq:targ}
  U_T^{(s)} \in \{\I\otimes\I, \Had\otimes\I, \T\otimes\I, 
                  \I\otimes\Had, \I\otimes\T, \CNOT\},
\end{equation*}
where $\I$ is identity operator on a single two-level subspace (qubit),
$\T=\exp(-i\pi/8\sz)$ is a $\pi/8$ phase gate, 
\begin{equation*}
 \Had = \frac{1}{\sqrt{2}}\begin{pmatrix}1 &-1\\ 1 & 1 \end{pmatrix},
 \quad
 \CNOT= e^{-i\pi/4}\begin{pmatrix} \I & 0 \\ 0 & \X \end{pmatrix}.
\end{equation*} 
Fig.~\ref{fig4} shows a possible solution switching sequences, which was 
obtained using an optimisation routine that took into account only the 
number of switches and overall gate operation times required to achieve 
the target fidelity of $99.99\%$.  In practice, it may be desirable to 
consider other factors such as penalties for rapid switching, etc to 
obtain the most robust and experimentally feasible solutions.

\section{Conclusion}

We have shown that there are different paradigms for quantum control
from open-loop Hamiltonian and reservoir engineering, to
measurement-based feedback control and coherent (quantum) feedback
using quantum systems as controllers, all of which have important 
applications, and different paradigms may be combined, as in the 
case of indirect control of one quantum system such as nuclear spins
by another such as electron spins, the latter being itself subject
to open-loop or measurement-based feedback control via external fields.
An important objective for control engineering is optimal control.
In the context of quantum systems the term has often come to mean
control via shaped pulses designed to optimise certain quantities 
such as the expectation value of an observable, yield of a chemical
reaction, or fidelity of a quantum process.  Especially in the latter
type of applications, time-optimality has often been considered to be
of paramount importance beside ensuring high fidelity.  However, we
have seen that there are other paradigms for optimal control.  In
particular in the context of quantum information processing, but not
necessarily limited to this application, fault-tolerance is an 
important consideration for optimal control that has received little 
attention so far.  Another important paradigm for optimisation is 
simplifying architectures with a view to eliminating unnecessary 
actuators and replacing complicated actuators by simple switches, for 
instance, while still maintaining the ability to effectively control 
the system.  Naturally, there are many open questions, in particular 
in novel areas such as optimal control and fault-tolerant circuit 
design, or minimalist control using the simplest possible actuators, 
ranging from the compatibility of optimal control and fault-tolerant
networks to the optimal switching sequences for binary switch actuators.

\section*{Acknowledgement}

SGS is an \EPSRC\ Advanced Research Fellow and acknowledges partial
support from the \EPSRC-funded \QIPIRC\ and Hitachi.  GK was supported
during the course of this work by the University of Melbourne's Centre 
for Quantum Computer Technology.  
 

\bibliographystyle{IEEEtran}
\bibliography{/home/sonia/archive/bibliography/References,IEEEabrv}
\end{document}